\documentclass[10pt, conference, compsocconf]{IEEEtran}
\ifCLASSINFOpdf
\else
\fi
\usepackage[cmex10]{amsmath}
\newtheorem{theo}{Theorem}
\newtheorem{asum}{Assumption}
\newtheorem{defi}{Definition}
\newtheorem{nota}{Notation}
\newtheorem{algo}{Algorithm}
\newtheorem{exam}{Example}

\usepackage{algorithmic}
\begin{document}
%
\title{Farsighted Collusion in Stable Marriage Problem}


\author{\IEEEauthorblockN{Mircea-Adrian Digulescu}
\IEEEauthorblockA{Computer Science Department\\
Faculty of Mathematics and Computer Science, University of Bucharest\\
Bucharest, Romania\\
mircea.digulescu@fmi.unibuc.ro\\
mircea.digulescu@gmail.com}
}


%


\maketitle

\begin{abstract}
The Stable Marriage Problem, as proposed by Gale and Shapley, considers producing a bipartite matching between two equally sized sets of boys (proposers) and respectively girls (acceptors), each member having a total preference order over the other set, such that the outcome is stable. In this paper we consider the Game directly induced by this problem and analyze the case when proposers collude. We present a linear time method for determining the unique optimal collusion matching which is farsightedly stable, under the following assumptions: (i) the sole utility in the Game is the rank of the match in own preference list (in particular, proposers are indifferent as to how other proposers fare); (ii) proposers make proposals iff farsightedly such plays would strictly improve their own outcome (thus proposers cooperate by refraining from making proposals which can only harm others, but not strictly help them; also, they cannot make concessions to others which harm themselves). We argue that this optimal outcome is actually stronger than a Strong Nash Equilibrium - no alternative feasible coalition exists which can offer at least one member a strictly better outcome under these assumptions. We also show why some prior results pertaining to collusion of proposers do not always yield a realistic outcome. \\
The results in this paper are an independent rediscovery of results by Jun Wako (2010), derived in a simpler fashion and phrased such that less jargon is employed.

\end{abstract}


%
\IEEEpeerreviewmaketitle

\section{Introduction}
The stable marriage problem was introduced to literature largely by Gale and Shapley in [1] where they provided an $O(n^{2})$ algorithm which produces a stable matching between the two sets. A stable matching in their view is one such where there exists no boy and no girl which both prefer each other over their respective partners in the matching. This can be viewed as strict stability or short-sighted stability. Once such a matching is attained it obviously cannot be broken since any individual proposal is doomed to failure: no boy (proposer) would ever be interested to make a proposal to any girl who would accept him over her then-current partner.

In this paper we take a rather different approach to stability. Namely we consider that a matching is stable if it is\textit{farsightedly stable}. That is, if no boy (or set of boys for that matter) can make a proposal or set of proposals (after dumping their original partners in the matching) such that, \textit{given the way in which all the boys will play as a result} – the subsequent proposals made - his (their) final outcome would strictly improve. The concept for farsighted stability implies that a boy will not make any moves which \textit{ultimately} do not strictly benefit him. So a matching that is farsightedly stable is not necessarily \textit{strictly stable} (Gale-Shapley stable). However it nevertheless represents a final (stable) outcome of the game so long as all actors understand that they cannot improve by any strategy.

\subsection{Prior work}
The stable marriage problem was introduced to literature largely by Gale and Shapley in \cite{cit01} where they provided an $O(n^2)$ algorithm which produces a stable matching between the two groups. A stable matching is one such that there exists no boy and no girl which both prefer each other over their respective partners in the matching. Robert W. Irving studied in \cite{cit02} the case where the preferences lists allow indifference between different partners and provided several algorithms for determining stable matchings (if they exist) in such cases. A few fundamental issues concerning strategic play were studied in \cite{cit03} by Dubins and Freedman where they showed that no coalition of boys can improve the outcome for all of them, by lying about their preferences.  In a sequel paper \cite{cit04}, Gale and Sotomayor analyzed the case with lying by the girls. Further analysis of strategic play by girls has been performed by Deng, Shen and Tang in \cite{cit18}. In \cite{cit05} the authors concerned themselves with lying by the boys in matching markets where one side only has a constant number of preferences and also acknowledged that ``no matching mechanism based on a stable marriage algorithm can guarantee truthfulness as a dominant strategy for participants". Another prior paper concerning itself with misrepresentation is \cite{cit06}. Some variations to the stable marriage problem concerning the simultaneous introduction of ties and seeking a more balanced matching (favoring the girls more than in Gale-Shapley) have been shown in \cite{cit07} to be hard (with regard to NP completeness). A series of existential results concerning strategic play in stable marriage problem and some of its variations have been presented by Roth in \cite{cit08}.
In \cite{cit09} the authors analyzed strategic play by girls and offered several results. Finally, in \cite{cit10}, Huang analyzed cases of strategic plays by boys consisting of collusion in order to achieve a better outcome for some (not all) of them. The novelty was that such outcomes are not necessarily stable. They are however no worse than Gale-Shapley and are also on the Pareto frontier of outcomes no worse than Gale-Shapley. This prompted Huang to imply that such outcome is the best possible for boys. His approach consisted of improving a Gale-Shapley matching by discovering and materializing ``trading cycles", using the top-trading-method described in \cite{cit11}. Using Huang's work, in \cite{cit19}, Aksoy, Azzam, Coppersmith, Glass, Karaali, Zhao and Zhu discuss marriage problem allocation striving to balance stability with efficiency. Since our paper proposes an realistic matching unlike \cite{cit10}, it is relevant to the discussion in \cite{cit19}.
A brief survey of results concerning stable marriage problem was published in 2008, in \cite{cit12} by Iwama and Miyazaki. A general and very relevant situation where the actual utility gained by men in a matching can be modeled in terms of transferable utility (e.g. monetary value) with side-payments allowed has been studied by Rahul Jain in \cite{cit13}.
Furthermore, just as we, the Author, were about to publish this paper for peer review, we discovered that Jun Wako \cite{cit21} had also discovered esentially precisely the results of this paper, relying on concepts such as von Neumann-Morgenstern stable sets, core and others. We present a simpler and much more direct approach, requiring very few if any theoretical prerequisits.
The concept of farsighted stability has been employed by Klaus, Klijn and Walzl for the related room-mates matching problem in \cite{cit17}.

\subsection{Overview of this paper}
The main result in this paper is a method for obtaining the best possible farsightedly stable matching for some boy. We proceed to show that in fact there is a unique farsightedly stable matching which is best for any of the boys (so the individual best outcome for an arbitrary boy is obtained in this one). We start by providing an $O(n^{3})$ algorithm and then improve it to a $O(n^2)$ one.

The second result is that we argue why the matching produced by the top-trading-cycles method used by Huang in [10] is not realistic. We give an example to illustrate the difference between our method and [10] and discuss why the former produces a realistic outcome while the latter does not. In the terminology of [10], we claim that Huang’s result is sometimes unrealistic because it entails cooperation of more accomplices than needed, with some accomplices actually having the incentive to refuse cooperation since they can obtain a better outcome (become cabalists in terminology of [10]) by not cooperating.

Finally we illustrate by example a situation where boys can lie about their preference lists (to other boys, before the actual game play) in order to alter the optimal farsighted stable outcome to their favor. This situation is not possible under the assumptions of the present paper – since lying actually implies a boy can make a proposal which does not improve his own outcome but to the contrary – it makes it worse (with regard to his truthful preference list), however it is worth noting nevertheless. 

\subsection{Motivation}
This paper is motivated primarily by effect of our results on the numerous economic situations which can be formulated as instances of the stable marriage problem. There are many examples in literature, such as Renter-Landlord, Contractor – Project, Student – Program placement, Ads placement and Wireless Communications [14] where our results can apply. For almost all such applications strategic play can play an important role. As such, we consider important the limitations of [10] (which is, to the best of our knowledge, the current state of the art in this regard) are noted and the qualitative improvements we make are considered.

\section{Preliminaries}
We consider the Game directly induced by the Stable Marriage Problem as follows: there is a set of players (the proposers, the boys) and a set of acceptors. Players can propose to acceptors during the game play. An acceptor (a girl) always accepts (or keeps) the best proposal and dumps (refuses) any lower one. As such, in the scope of this paper, we ignore any potential strategic behavior by acceptors. They are considered robotic elements with no actual decisions. The utility in the game for each player is the rank of his final match (the one after the game ends) – the lower this rank (the more preferred the match) the strictly higher the utility. The game is considered to end when no proposer (boy) can take any action \textit{which would \underline{farsightedly} \textbf{strictly}} improve his outcome.

Please note that the final condition implies the game can end before a strictly stable (Gale-Shapley stable) matching is attained. Also note that for the Game to end, no set of the boys can have a strategy available to them so that, given the strategy of the other players, the former could hope to improve.

Speculative play – when a boy proposes to some girl and finally dumps her (during the course of the game, not during preliminary “negotiations”) is actually irrelevant once the game has actually begun, under our assumption set: the boy will only be interested to dump if, by dumping he can improve his own final outcome. And if he has a strategy of improving this final outcome – he could just as well not have proposed to her in the first place but instead chosen the alternative strategy directly.

\subsection{Assumptions}
When considering coalitions we need to distinguish between feasible and unfeasible ones. What makes a coalition unfeasible? While it can be philosophically difficult to describe objectively what makes a coalition unfeasible in this context, it can be done. Intuitively we will say that a coalition is unfeasible if it requires that a boy gets a \textit{strictly} worse outcome than he can get if he did not part-take in the coalition (require him to make a concession). However, this strictly worse outcome is a function of the behavior and predilections of all the other players in the game. Thus, in order to be able to answer this question we need to examine these behaviors also. Note that while girls are not considered players, having no decision points, their preference lists \textit{are very relevant} to the outcome of the collusion between boys. A boy can end up matched with a girl $g$ by persuading all boys who are better preferred than him by $g$ to never propose to her in the game. However, he does not need to care about the rest – their cooperation is not required.

In the scope of this paper we introduce two natural assumptions in order to be able to reason about how boys behave. These assumptions do not limit generality significantly and they are strictly upheld in most situations occurring in practice. They are also implicitly made in most of the existing literature (although interesting situations can arise when they are relaxed). All in all, the case where they hold is worth investigating in its own right.

\begin{asum}
The sole utility boys seek to optimize in the game is the rank on their preference list of the match they get.
\end{asum}
Thus, they will always strictly prefer an outcome that offers them individually a better (final) match. Thus, they will never accept to part-take in a coalition which offers them a worse outcome than one \textit{guaranteed} by a different coalition (e.g. consisting of only themselves). Also, they will be fully indifferent as to how the other boys fair so long as it does not affect their own matching. In particular they will not “take sides” and will not care which one of the other boys fairs better than the other. It could, of course, be interesting to examine (especially in the context of repeat games), behaviors of boys which are friends or allies and favor one-another. This potentially interesting situation is outside the scope of this paper. 

\begin{asum}
Proposals are made \textbf{iff} farsightedly they improve the outcome for the proposer.
We require of players who can make such a proposal that, given how the other players play as a result, they would get a strictly better outcome at the end of the Game, that they indeed make one such proposal. Also we require that a player who does not have such a proposal available to him, that he does not make any proposals (even if he can have a proposal which would improve his outcome, but only temporarily – before the other players finish making their plays).
\end{asum}

Assumption 2 actually excludes (i) concessions and (ii) threats in negotiation of the coalition outcome. Note that while concessions are typically excluded as irrational behavior anyway, in practice sometimes players may not know whether they are making a concession or not due to insufficient theoretical knowledge of the Stable Marriage topic. For example, some required accomplices for the coalition resulting from Huang’s approach [10] might have accepted to join such a coalition (before reading our paper) even though it did not offer them the best outcome they could get. When it comes to threats, Assumption 2 essentially excludes “suicidal threats” – when a player says (before the game is played!) something like “if you don’t  agree to give me my desired girl, I will make sure you don’t get your desired girl either, \textit{even if it means I get a worse outcome then I could have gotten}.” Such threat making may not necessarily be irrational – the threat may be feasible and it might work, thus resulting in the threat maker being persuasive. Threat making greatly complicates game analysis and is also risky for the threat-maker: the instance he proposed to a worse girl than he is guaranteed to get under Assumptions 1 and 2, the others (in particular just the sole boy from whom he takes this girl) can punish him by letting him matched with this undesired choice. This in fact leads to situations of a Game of Ultimatum. Such situations, while they can be very interesting and generate surprising outcomes, are outside the scope of this paper. They will be discussed in a subsequent one. 

\subsection{Conventions}

\begin{nota}
We use the following notation to denote a play at some stage of the Stable Marriage Game. $b_x \rightarrow g_y(b_t)|b_z \rightarrow  ... \rightarrow g_z$, with the meaning that $b_x$ proposed to $g_y$ and as a result $b_z$ (we can have $b_z$ = $b_x$) was “expelled” from $g_y$ (losing to $b_t$) and went on to propose to some other girl and so on until some boy proposed to $g_z$, who was unmatched, ending the play. We call each portion separated by | an element of the play. We can omit the parenthesis (like $(b_t)$) since it is obvious who was the victor in that element by analyzing the next one.
\end{nota}

\begin{nota}
We use the following notation to denote a cooperative switching of partners: $g_y|b_x \rightarrow fz|bz \rightarrow ... g_y$, with the meaning that boy $b_x$ renounces his then-current partner $g_y$ and successfully proposes to girl $g_z$ whom $b_z$ renounced and so on, until a boy proposes to the original (now unmatched) girl $g_y$.
\end{nota}

Note that a trading cycle may not always be feasible under Assumption 2 (lead to a farsighted improvement). Unlike in other resource allocation problems where the resources are fully passive, the fact the girls of the Stable Marriage problems keep the best proposal they get affects the relative power of the boys. They are not always free to trade their partners as they choose since once traded, other boys (who would have been rejected under the initial matching) could now successfully propose to the girls of the trading cycle, effectively vetoing it. This is something that was not fully properly considered in [10]. Note however, that, under Assumption 2, boys can (and must) veto only if this results in them farsightedly improving their own outcome; not for revenge or other purposes.

\begin{defi}
Temperature of girls. We call the rank in her preference list of the then-current partner of a girl during the course of the game as the temperature of said girl. 
\end{defi}

Note that since girls are considered robotic in the scope of this paper, the temperature of any girl is non-decreasing throughout the Game-Play: girls get better and better partners, never worse. 

\section{FARSIGHTED STABLE MATCHING }
We now proceed to examine how boys can collude to improve their individual outcome in the Stable Marriage Game.

It is known that Gale-Shapley’s algorithm offers the unique man-optimal strictly stable matching ([1],[10]). No other strictly stable matching exists which offers any of the boys a better outcome. We present a similar algorithm which generates the unique man-optimal farsightedly stable matching. 

We call a matching \textit{farsightedly stable} \textbf{iff} under Assumptions 1 and 2, there are no more allowed plays by any of the boys. We can allow boys to dump girls from this matching (including all of them all girls) but we do require in the context of farsightedness that every boy eventually proposes to some girl (no boy can remain unmatched). Note that if we had relaxed Assumption 2 to allow moves which farsightedly produce no \textit{strict} improvement, then the Gale-Shapley matching would be farsightedly stable.

When investigating the coalitions in which a boy could part take, we first ask ourselves: what is the worse outcome that could happen for him?

\begin{theo}
Under Assumptions 1 and 2, every boy in a Stable Marriage Game gets no worse an outcome than in Gale-Shapley. Regardless what coalitions form and no matter what their strategies are, every boy in a Stable Marriage Game has a strategy to ensure that he gets a partner no-worse than Gale-Shapley, so long as Assumptions 1 and 2 are upheld.
\end{theo}
\begin{IEEEproof}
By way of contradiction we shall show that no worse outcome of the Game can occur. Let us consider any potential final outcome of the game where some boy $b$ gets a worst match than in Gale Shapley. Say boy $b$’s Gale-Shapley partner was taken up by another boy, $b_1$ in this outcome. By induction we obtain a ring of boys, $b, b_1, b_2, ..., b_k$, all of whom occupy the Gale-Shapley partner of their left neighbour and have their own Gale-Shapley partner occupied by their right one (wrapping around if necessary). Clearly $b$ prefers if he were coupled with his Gale Shapley partner (contradiction hypothesis). But $b_1$ also prefers the hypothetical outcome worse than if he were coupled with his own Gale Shapley partner (otherwise he would have proposed there in the course of Gale Shapley algorithm, meaning that would no longer have been the final partner for b). Similarly so do $b_2,...,b_k$. But then a trading cycle forms where each boy becomes coupled with the partner of his right neighbor (wrapping around). Since this trading cycle would offer strictly better outcomes to all players on it, by Assumptions 1 and 2, they are all intrested in its materialization. Could any of the other players veto the new arrangement? Not really. Say another boy $b_z$ is interest to successfully propose to the match of one of the boys on the cycle (after materialization) over his own. But the boy on the cycle is coupled with his Gale-Shapley partner; as such, he is best prefered there over all other boys who get a Gale-Shapley or better outcome (and thus never propose to less preferred choices). It follows that $b_z$ also got a worse match than in Gale-Shapley. Furthermore, $b_z$ prefers his own Gale-Shapley over both his then-current one and over the match of any boy in the trading cycle where he could successfully propose. As such, $b_z$ is either part of the original trading cycle, or another such cycle emerges. Inductively, the set of such cycles is a farsighted improvement for all boys involved. As such, it follows that the Game had not ended, contradicting the initial assumption that the outcome was final. Note that the actual move to be played may not be materializing these cycles (some boys might be able to get even better) – but if no other move exists (which farsightedly improves things for someone), then this one is definitely valid.
\end{IEEEproof}

The theorem implies that no boy will support a coalition offering him worse than Gale-Shapley.

In every Stable Marriage Game, there exists at least one player who cannot get better than Gale Shapley either. It is very easy to find one: consider the player who made \textit{the last proposal} in the course of Gale Shapley algorithm: he gets a girl to whom nobody else ever proposed and also all his better preferred choices are taken by people who are better preferred than him there. Such a player is a “hopeless man” in the terminology of [10]. We show that he is also hopeless considering farsighted stability.

\begin{theo}
The sole proposer to some girl in Gale Shapley is a hopeless man, under Assumptions 1 and 2. There exists no outcome where such a boy can farsightedly get a strictly better match. 
\end{theo}
\begin{IEEEproof}
Similar results have been proven throughout literature. For completeness we include our proof also.
Proof proceeds by contradiction. Say that such a boy $b$ would get a better partner than Gale Shapley, $g_1$. But $g_1$ is the Gale Shapley partner of some other boy, $b_1$. For the outcome to be final, it follows that $b_1$ must get a strictly better outcome than $g_1$ (since, by Theorem 1, he can get no worse). Again, a ring of boys $b, b_1,..., b_k$ forms, where each  boy takes the Gale Shapley partner of the boy to the right (wrapping around). Furthermore, all boys can force their left neighbor out. All boys must prefer this outcome over the Gale-Shapley one. But $b_k$ gets as partner the Gale-Shapley partner of $b$ which is strictly worse for him than his own Gale-Shapley partner (otherwise he would have proposed to her before his final match in the course of that algorithm). By Assumption 2, $b_k$ cannot support such a coalition as per Theorem 1 he is guaranteed to get at least his Gale-Shapley partner if not better. In particular $b_k$ can simply knock out his left neighbor in the ring who can also proceed to knock out his left one and so on. 
\end{IEEEproof}

By Theorems 1 and 2 we immediately derive the optimal farsighted outcome for a hopeless man: it is his Gale-Shapley match. By Assumption 2, if he gets this outcome he \textit{must not} make any proposals. But there are several ways in which this hopeless man can attain it: the worse (for the others) is if he proposed in order of preference to all girls until getting this outcome. However, by Assumption 1, this play would be just as good to him as the play where he makes a single proposal to this optimal partner. The hopeless man has no incentive (in the context of a game) to help others, but no disincentive either! Furthermore, under Assumption 2, once he realizes his optimal outcome he must not make any more plays. In particular, if the boys who are (temporarily) coupled with some other girls all decided to dump their partners and “renegotiate” the situation, the hopeless man would be (by Assumption 2) required to refrain from making any plays: no matter what he did, farsightedly there is still no way he could end up strictly better off (by Theorem 2). 

A hopeless man effectively has a very simple strategy which guarantees he will get his optimal outcome: play directly there. If he does so, given Theorem 6 no boy will challenge him there. Furthermore, he can help the others by not having uselessly raised the temperature of some girls during game play. In practice he could still veto some coalitions at no cost to him (change his strategy to disallow certain outcomes) - but by Assumption 1 this cannot benefit him. While in practice he could, at no cost, very well be spiteful or vengeful in case he wanted to help a friend and the others refused, within the scope of this paper this is explicitly disallowed by Assumptions 1 and 2. In some practical implementations (for example Student – Medical Program matching), players are not allowed to dump partners. However we can make our Assumptions 1 and 2 hold in this case simply by requiring that a hopeless man plays to his optimal outcome directly.

Since a hopeless man ends up coupled with his Gale Shapley partner (whom no one else wants), we can simply eliminate him from the set of players and then repeat the procedure until no more players are left.
\begin{algo}
\end{algo}
\begin{algorithmic}[1]
\STATE Consider the set of boys $B$ and of girls $G$, with preference lists $PB$ and $PG$ respectively.
\STATE Let $(b_1,g_1)$ be the final proposal made in running Gale Shapley algorithm for instance $(B, G, PB, PG)$.
\STATE Match $b_1$ to $g_1$ permanently.
\STATE Remove $b_1$ from $B$ and $g_1$ from $G$.
\STATE If the set of boys is not empty, go to Step 2. 
\end{algorithmic}
\textit{Correctness}: Would it ever make sense (farsightedly) for a boy $b$ still in play to successfully propose to some girl who has been eliminated at some prior iteration of Step 4? Not really. Consider for the sake of contradiction the earliest boy $b_1$ who was eliminated at Step 4, who would, in this hypothetical scenario by kicked out of his girl $g_1$ by some boy not eliminated before. Since there are no proposals made to girls eliminated at prior stages, those girls (and their partner boys) can be fully disregarded form the subsequent problem instances. But by Theorems 1 and 2 boy $b_1$’s match is precisely $g_1$. As such, any resulting matching will not be final (since in the final farsighted stable matching boy $b_1$ remains coupled with precisely $g_1$). The move by boy $b$ cannot farsightedly improve his outcome and is thus disallowed by Assumption 2. 

\textit{Complexity Analysis}: For Stable Marriage problem with $n$ boys, the algorithm essentially consists of $n$ runs of the Gale-Shapley algorithm, the first on a problem with all $n$ boys, then on one with $n-1$ boys, then on one with $n-2$ boys and so on until there is only one boy remaining. The running time for Gale-Shapley is $O(n^{2})$ for $n$ boys, thus the total running time is $O(1^{2} + 2^{2} + ... + n^{2}) = O(n^{3})$. The memory consumed is also dominated by Gale-Shapley (but can be reused by different iterations of Step 2) and is thus $O(n^{2})$. 

\textit{Uniqueness}: Note that the order in which the boys are presented to Step 2 of Algorithm 1 can affect the choice of hopeless man in Step 3. However, if a man was hopeless before the elimination of another hopeless man, he is still hopeless afterwards: In the proof of Theorem 2 such a hopeless man could not have been part of the ring $b, b_1,...,b_k$ of boys in any other capacity than b (since his left neighbor cannot prefer his Gale Shapley partner – to whom he never proposes – over his own). So eliminating some other hopeless man leaves the ring constructed in the proof applicable also on the outstanding problem instance to all hopeless men not eliminated. Furthermore, the Gale-Shapley partner of these hopeless men does not change: by the prior argument it cannot get better, the elimination of a boy cannot degrade it, and the elimination of a girl irrelevant to all boys cannot degrade it either. As such, the order in which hopeless men are eliminated by Step 4 is irrelevant to the final outcome, which is thus unique for a particular Stable Marriage problem instance.

The significance of the Uniqueness of the farsighted stable matching produced by Algorithm 1 cannot be understated. It follows that Algorithm 8 produces \textbf{\textit{the unique man-optimal farsightedly stable matching}}. In particular there exists no farsightedly stable matching which gives any of the players a strictly better outcome (degrading or not some of the rest). As such, this collusion outcome is even stronger than a Strong Nash Equilibrium (which requires only there exists no alternative coalition improving \textit{all} of its members’ outcomes, where as we show that there exists none which improves \textit{any} of its members). In fact there exists no other farsightedly stable matching, under the natural Assumptions 1 and 2. 

\section{LINEAR ALGORITHM }
The idea behind the linear time algorithm we present in this section is to avoid having to do repeat runs of Gale-Shapley in Step 2 of Algorithm 1, but instead somehow keep and update some state which allows us to find a hopeless man directly.

Consider a Gale-Shapley implementation where boys are introduced to the problem one by one, in some order. When a boy is introduced, a round of play happens starting with him proposing to his best preferred girl and either kicking someone out or continuing to his next preference. The then-uncoupled boy (who may be the same as before the proposal) proposes to his next preference and so on until finally some boy (not necessarily the one initiating in the round) proposes to some unmatched girl, ending the round. As a result of a round, the matching for some (or all) of the boys can change.

Consider the permutation of the boys consisting of the reverse order of their elimination in Step 4 of Algorithm 1. If we happened to get lucky and run Gale-Shapley by introducing the boys in this order, after the first run, updating the state would be very easy: we would simply remove the last play (necessarily made by a then-hopeless man), as the remaining rounds would form precisely a Gale-Shapley iteration for the outstanding boys.

While we cannot know the correct order in advance, we are able to reconstruct it incrementally – on the fly – as we eliminate hopeless men. We do this by determining what the last round of play would have been if the boys had been presented in the ideal order.

Note that as boys are eliminated from a Stable Marriage problem instance, the set of proposals made under Gale-Shapley by the outstanding players either contracts or stays the same. This follows trivially if we consider the eliminated boy as initiating the last round in a Gale-Shapley implementation.

As such, if a hopeless man was to have been presented last to Gale-Shapley, then his round would still be a play consisting of a subset of all the proposals in the game.

The algorithm proceeds by essentially identifying the precise set of proposals which are part of such a last round, should the hopeless man have been presented last to Gale-Shapley algorithm implementation and then eliminates them. Equivalently, it determines the Gale-Shapley matching after the elimination of the hopeless man and all his proposals. It does so by repeatedly eliminating trading cycles which cannot be legitimately vetoed at the time.

\begin{algo}
The algorithm consists of three routines and uses some global variables, as follows. \\
\underline{Input}: Let $n$ denote the number of boys (and of girls). \\
Let $PB_{b,i}$ and $PG_{g,i}$ be $n$ by $n$ matrices representing the preference lists of boys and girls respectively. We require that the sort order of preferences for the boys is best prefered first, while for the girls is best prefered last. \\
\underline{State}: Let $EXISTS_{b,g}$ be a $n$ by $n$ matrix of Boolean values initially set to \textbf{false}. It is used to represent if a proposal from a boy to a girl exists. \\
Let $Index_b$, be an $n$-sized integer vector representing the index in his preference list of the worse proposal by a boy still in play. This must be initialized to $-1$. \\
Let $Top_g$, be an $n$-sized integer vector representing the index of the current match of a girl in her preference list. This must be initialized to $-1$. \\
Let $Second_g$, be an $n$-sized integer vector representing the index of the immediately less preferred choice for a girl $g$ after top one, from all men who ever proposed to her. \\
Let $NumProposals$, be an $n$-sized integer vector used to maintain the current number of proposals still in play made to some girl $g$. \\
\underline{Helpers}: Let $seen_b$, be an n-sized integer vector used in depth first walk, holding values only $0$, $1$ or $2$. \\
\underline{Output}: Let $M_b$ be an n-sized integer vector. It will hold the final partner for any boy$b$ after $FindMatching()$ is called.
\end{algo}

\begin{algorithmic}[1]
\STATE \textbf{METHOD $FindMatching()$}
\STATE Run of Gale Shapley algorithm for $(B, G, PB, PG)$.
\FORALL {proposals $(b, g)$ made in line 2, in that order}
	\STATE $EXISTS_{b,g} \leftarrow true$
	\STATE Increment $Index_b$
	\IF{proposal $(b,g)$ is successful}
		\STATE $Second_g \leftarrow Top_g$
		\REPEAT
			\STATE Increment $Top_g$
		\UNTIL{$PG_{g,Top_g} = b$}
	\ENDIF
\ENDFOR
\FOR{exactly $n$ iterations}
	\STATE $g_1 \leftarrow$ some arbitrary girl with $NumProposals_{g_1} = 1$
	\STATE  $b_1 \leftarrow PG_{g_1,Top_{g_1}}$
	\STATE $M_{b_1} \leftarrow g_1$ and remove $b_1$ from play.
	\STATE $EliminateProposalsByBoy(b_1,-1)$
	\STATE Initialize $seen_b \leftarrow 0$ for all boys $b$ still in play.
	\FORALL{boys $b$ still in play} 
		\WHILE {$seen_b = false$}
		\STATE $FindAndEliminateTradingCycles(b)$
		\ENDWHILE
	\ENDFOR
\ENDFOR
\RETURN $M$
\STATE \textbf{ENDMETHOD}
\STATE \textbf{METHOD $EliminateProposalsByBoy(b, g_0)$}
\WHILE {$Index_b \geq 0$}
	\STATE $g \leftarrow PB_{b,Index_b}$
	\IF{$g = g_0$}
		\STATE \textbf{break}
	\ENDIF
	\STATE $EXISTS_{b, g} \leftarrow \FALSE$
	\IF{$Top_g = b$}
		\STATE $Top_g \leftarrow Second_g$
		\STATE Decrement $Second_g$
	\ENDIF
	\WHILE{$Second_g \geq 0$ and $EXISTS_{PG_{g,Second_g},g} =$ \FALSE}
		\STATE Decrement $Second_g$
	\ENDWHILE
	\STATE Decrement $Index_b$
	\STATE Decrement $NumProposals_g$
\ENDWHILE
\STATE \textbf{ENDMETHOD}
\STATE \textbf{METHOD $FindAndEliminateTradingCycles(b)$}
\STATE $seen_b \leftarrow 1$
\STATE $g \leftarrow PB_{b,Index_b}$
\STATE $b_2 \leftarrow PG_{g,Second_g}$
\IF{$b_2 = -1$ or $seen_{b_2} == 2$}
	\STATE $seen_b \leftarrow 2$
	\RETURN $(-2,-2)$ \COMMENT{Dead end}
\ENDIF
\IF{$seen_{b_2} = 1$}
	\STATE $seen_b \leftarrow 0$ \COMMENT{Cycle found: mark $b$ unvisited}
	\RETURN $(b_2,g)$ \COMMENT{first boy in cycle and his new match}
\ENDIF
\STATE $(head, g_2) \leftarrow FindAndEliminateTradingCycles(b_2)$
\IF{$head = -2$}
	\STATE \textbf{goto} line 50 \COMMENT {Dead end}
\ENDIF
\IF{$head = -1$}
	\STATE \textbf{goto} line 46 \COMMENT {Cycle was eliminated, so repeat}
\ENDIF
\STATE $EliminateProposalsByBoy(b_2, g)$ \COMMENT{Give $b_2$ his new match, namely the girl currently held by $b$}
\STATE $seen_b \leftarrow 0$ \COMMENT{$b$ is part of the cycle, so mark him unvisited}
\IF{$head = b$}
	\STATE $EliminateProposalsByBoy(b, g_2)$ \COMMENT{Reached the head of the cycle}
	\RETURN $(-1,-1)$ \COMMENT{Tell the parent to repeat}
\ENDIF
\RETURN $(head, g_2)$
\end{algorithmic}

\textit{Correctness}: $EliminateProposalsByBoy(b,g)$ eliminates all proposals by boy $b$ to all girls less preferred than $g$, and updates the state accordingly. It is trivial to note that all State variables correctly maintain their semantics after a completed call to this method. Note that $Index$ might hold duplicate values sometimes (thus not representing a valid bipartite matching). However this is not of concern since it is never used with that semantic while this occurs. 

$FindAndEliminateTradingCycles(b)$ does a Depth First walk, starting at $b$, of a specially constructed directed graph over the boys, in the hopes of finding a trading cycle. This directed graph has an edge from each boy $b$, to the second best preference of the girl at $Index_b$, if such exists. This is actually the second preference of the girl who prefers $b$ best from all her proposers and whom $b$ prefers least from all his proposals, except while a trading cycle is materialized in lines 64-70.

Like with all Depth First walks, it could either reach a dead-end (either directly or by hitting a priory fully explored node - cross edge) or find a cycle. If it finds a cycle – we shall show it is an impossible to veto trading cycle -, it materializes it and then continues the search from right before it entered it (from the \textit{parent}). Note the fact this special graph has for each node at most one outgoing edge. When a cycle is materialized, the graph changes - but only slightly. Only the nodes on it plus the parent can have their either outgoing edge changed (or removed). This change can happen as a result of a boy $b$ having his $Index_b$ changed in lines 64 or 67 and thus potentially changing the boy who is immediately better preferred than him at that girl. Note that all such girls were priory occupied by boys also on the cycle. It follows that no fully explored node (status 2) can have an edge to any of the boys on the cycle – if it did so after the materialization of the exchange, it must have pointed to some boy on the cycle prior to this also, which means the cycle would have been explored earlier in the Depth First walk. Since each node has at most one outgoing edge it means no partially explored nodes (status 1) can point to any of the boys on the cycle even after its materialization, except potentially the parent. As such, after the cycle is materialized, the Deapth First walk can safely continue on the altered graph from the parent (which is re-explored thanks to lines 61-63): all explored or partially explored nodes and edges are the same.

Note the following invariants pertaining to how the graph is constructed and maintained. If there is an edge from $b_1$ to $b_2$, the following are true: (i) $b_1$ is best preferred at his match; (ii) $b_2$ prefers $b_1$’s match over his own and (iii) $b_2$ is second preferred at $b_1$’s match after $b_1$. Note that all these hold before and after a cycle is materialized in lines 64-70: boys on the cycle become matched to what is now a girl who prefers them best – thanks to invariant (iii). Invariant (ii) holds due to the elimination of proposals by a boy which are worse than his current match, in lines 64 or 67. Initially it holds due to the nature of Gale-Shapley algorithm: proposals are made by boys in order of preference stopping at their final partner. Invariant (iii) holds by the very construction of an edge in line 48.

Thanks to invariant (ii), a cycle found in Line 53 is a trading cycle. Thanks to invariant (i) it is also impossible to veto legitimately: any boy outside the materialized cycle will either have a better partner than a girl $g$ on the cycle, or be less preferred by $g$ than her current match (both before and after the trading cycle is materialized).

$FindMatching()$ repeatedly finds a hopeless man, eliminates his proposals and then repeatedly finds and eliminates trading cycles \textit{which are impossible to veto} until there are no more. We now show that the resulting matching, represented by the state variables is actually the Gale-Shapley matching for set of outstanding boys after each hopeless man had been eliminated. Proof is by contradiction.

Assume for the sake of contradiction there exists a proposal $(b_1,g_1)$ eliminated by line 19 which is actually part of the Gale-Shapley play for the outstanding boys. Take without loss of generality the first such proposal which is wrongly eliminated. It must have been eliminated either by the materialization of a trading cycle, or as a consequence of it become irrelevant. But once a trading cycle is materialized all boys are guaranteed to get at least as good a match. Thus clearly all proposals to less preferred girls must necessarily be removed once this happens. This means the erroneous proposal $(b_1, g_1)$ must have been one of those eliminated directly as part of the materializing the trading cycle. If any such proposal was to be kept, the cycle cannot materialize. But if $b_1$ does not leave $g_1$, can boy $b_2$ who was supposed to get $g_1$ as part of the trading cycle still improve over his then current $g_2$ (as part of a different trading cycle perhaps)? Not really: If $b_2$ was to improve, that means he would leave $g_2$. However, thanks to invariant (ii) b1 prefers $g_2$. Thanks to invariant (iii), if $b_2$ leaves $g_2$ (for whomever), then $b_1$ becomes best preferred there. As such it cannot be that both “proposal $(b_1,g_1)$ exists” and “proposal $(b_2,g_2)$ does not exist” are true, since Gale-Shapley produces a strictly stable matching. Thus, proposal $(b_2,g_2)$ must also remain. This means $b_2$ also does not improve over his partner before trading-cycle materialization. Inductively, none of the boys on it improve. Thanks to invariant (i), it follows that for all boys on that cycle have their partner before materialization as their final partner. This would mean the correct solution includes this trading cycle. However, the Gale-Shapley outcome is man-optimal over all strictly stable matchings [10]. As such, it cannot contain any trading cycle which cannot be legitimately vetoed, thus a contradiction, meaning all proposals $(b,g)$ eliminated by line 19 are not part of the Gale-Shapley run for the outstanding boys.

Now assume for the sake of contradiction that there is a proposal $(b_1,g_1)$ which is not eliminated although it should have been. Without loss of generality assume $(b_1,g_1)$ is the least preferred proposal by $b_1$ of those wrongly not eliminated. Consider the moment line 19 completes. By the first part of the argument, all proposals eliminated until then do not belong to the correct solution. Also, since it completes, it follows that there are no longer any trading cycles which cannot be vetoed legitimately. Thanks to invariant (i), we have that $b_1$ must be the best preferred choice of $g_1$. Furthermore, if $g_1$ is not $b_1$’s Gale-Shapley partner, it must be someone else’s, namely $b_2$’s. By the prior argument, all proposals by $b_2$ up until at least $g_1$ are still in the solution set produced, never having been wrongly eliminated. This means $b_1$ is not alone to propose to $g_1$: at least $b_2$ also did, if not others too. Now take $b_3$ to be the second best preference still in play of $g_1$. It could be the same as $b_2$ or different. Then let $g_2$ be $b_3$’s match in the solution proposed by the Algorithm. Clearly $g_2$ is different from $g_1$ since $b_1$ gests $g_1$, not $b_3$. Also clearly, $b_3$ prefers $g_1$ over $g_2$. If $b_3$ is not $b_2$, $b_3$’s proposal to g1 cannot exist in the correct solution, since this would imply $g_1$’s temperature is raised irremediably above $b_2$ who is supposed to be her Gale-Shapley partner. This means $b_3$ gets better than $g_1$ in the correct solution and in such a solution he never proposes to neither $g_1$ nor the worse preferred $g_2$. If $b_3$ is actually $b_2$, still the proposal by $b_3 (=b_2)$ to $g_2$ cannot exist since it would be beyond his Gale-Shapley partner. It does follow that $g_2$ is not $b_3$’s Gale-Shapley partner for sure (such a proposal necessarily exists in the correct solution). This means there are at least two proposals at $g_2$: one by $b_3$ and one by her Gale-Shapley partner. Note that there will be an edge in the graph from $b_1$ to $b_3$. Inductively we thus construct the path originating in $b_1$ and going along edges in the final graph. Since line 19 completed, this path cannot lead to a cycle (such a cycle would a trading cycle which cannot be vetoed). But also inductively no boy in the path is alone to propose at his current match. It follows there is an outgoing edge from each boy. Thus such a cycle must actually exist (not necessarily starting at $b_1$). We have obtained a contradiction implying that no proposal is wrongly kept either. 

Since no proposal is either wrongly kept or wrongly deleted, and since no new proposals can appear in the correct solutions, it follows that the output of line 19 actually performs a correct update of the matching given the elimination of a hopeless man. Thus, inductively, we have that the output produced by Algorithm 2 is the same as that of Algorithm 1. 

\textit{Complexity Analysis}: Consider the total number of iterations happening in line 39 in method $EliminateProposalsByBoy()$. Note that at every iteration, $Second_g$ is decreased for some girl $g$. By the loop condition, $Second_g$ cannot go below $-1$. $Second_g$ is initially the index of the second best proposal girl $g$ gets, which is at most $n$, having been initialized in line 7 of $FindMatching()$. It is never thereafter ever increased. As such there cannot be more iterations than $n * n = O(n^2)$ for all the $n$ girls. One execution takes $O(1)$ so the total amount of time taken by this line is $O(1)*O(n^2) = O(n^2)$. Now consider how many times lines 29-42 execute in total. Note that in line 41 $Index_b$ is decremented for some boy $b$ and by condition in line 28, it never goes below $-1$. $Index_b$ is initialized in line 5 by $FindMatching()$ to be precisely the number of proposals $b$ makes. It is never thereafter ever incremented. As such, the total number of lines 29-42 execute is no more than the total number of initial proposals by all boys, which is $O(n^2)$. An actual execution of these lines takes O(1), except the time taken by line 39 which is $O(n^2)$ in total. As such, the total time consumed by $EliminateProposalsByBoy()$ is $O(n^2)*O(1) + O(n^2) = O(n^2)$. 

Line 19 of $FindMatching()$ actually does a Depth First walk using $FindAndEliminateTradingCycles()$ on a graph with at most $n$ nodes and at most $n$ edges. This walk is somewhat complicated by the fact certain nodes are considered more than once as cycles are materialized. What is the total number of times line 46 of $FindAndEliminateTradingCycles()$ executes in total? It executes precisely one time for each boy $b$ – the last time – when it eventually returns $-2$ for each execution of line 19 in $FindMatching()$, plus one more time for every case when some $seen_b$ was set to 0 by lines 54 or 65 plus another more time for each case when line 62 calls for a repeat due to a cycle having been materialized. Note that every time line 54  or 65 happen, at least one proposal is eliminated from play by the subsequent application of Line 64 or 67 as the recurrence folds back. As such the total number of times $seen_b$ is set to 0, in lines 54 or 67 for all boys $b$ cannot be greater than the total number of initial proposals, which is $O(n^2)$. How many times does line 62 call for a repeat? Precisely every time when a trading cycle materialization completes in Line $68$. Since materializing any trading cycle causes at least 2 proposals to be removed, there cannot be more than half as many trading cycle materializations as there are initial proposals, which is $O(n^2)$. As such, the total number of times line 62 calls for a repeat is $O(n^2)$. Thus, the total number of times the body of $FindAndEliminateTradingCycles()$ executes is $n*O(n) + O(n^2) + O(n^2) = O(n^2)$. The running time of the body itself is $O(1)$ – it contains no loops – plus the time taken within calls to $EliminateProposalsByBoy()$. The latter time we showed to be $O(n^2)$. As such, the total running time of $FindAndEliminateTradingCycles()$ is $O(n^2)*O(1)+O(n^2) = O(n^2)$.

Line 14 of $FindMatching()$ is clearly $O(n)$ since all girls could be exhaustively tried (it could be made even faster but there is no need), as do lines 18 and 19, excluding the time taken by $FindAndEliminateTradingCycles()$ which is $O(n^2)$ in total. Lines 15-17 all take O(1) except time consumed by $EliminateProposalsByBoy()$ which is $O(n^2)$ in total. As such, line 13 total takes $n*[O(n)+O(1)] + O(n^2) + O(n^2) = O(n^2)$ for all the $n$ iterations. The total number of times line 9 executes for some girl $g$ cannot exceed the index of her best proposal, which is at most $n$. As such, for all $n$ girls, line 9 takes at most $n * n = O(n^2)$ time. Lines 4-11 take $O(1)$ each time they execute, plus $O(n^2)$ in total for line 9. They are executed for each proposal happening as part of Gale-Shapley algorithm, of which there can be at most $O(n^2)$. Thus, the total running time of line 3 is $O(n^2)*O(1) + O(n^2) = O(n^2)$. Line 2 takes $O(n^2)$ and line 25 takes $O(1)$. Thus $FindMatching()$'s total running time is $O(n^2) + O(n^2) + O(n^2) + O(1) = O(n^2)$.

We have thus shown that Algorithm 2 takes $O(n^2)$ running time in total. Given that there are $O(n^2)$ preference list items given as input, this is linear in the size of such input.

In terms of memory complexity, only $O(n^2)$ or $O(n)$ space is used by any of the constant number of State or Helper structures employed. The memory complexity of running Gale-Shapley once is also $O(n^2)$. As such, the total memory complexity of Algorithm 2 is $O(n^2)$ which again is linear in the size of the input.

We have implemented $FindAndEliminateTradingCycles()$ using recursion. This was done mainly for clarity and the method could easily be rewritten without it. The total number of simultaneous recursive calls is $O(n)$, since there are at most $n$ boys in total in the graph. This could become a problem in practice if $n$ is large enough to cause a stack overflow. 

\section{TOP TRADING CYCLES METHOD DOES NOT ALWAYS PRODUCE THE FARSIGHTEDLY STABLE MATCHING}
In [10], Huang proposed using the top-trading-cycles method known from housing allocation to improving the outcome for the boys in a Stable Marriage Game. We know show that this approach is not ideal, since it sometimes requires the existence an unfeasible coalition. We show this by means of counter example. Consider the following instance of the Stable Marriage game.

\begin{exam}
$\newline$
$B=\{b_1,\ b_2,\ b_3,\ b_4,\ b_5,\ b_6,\ b_7\}$
$\newline$
$G=\{g_1,\ g_2,\ g_3,\ g_4,\ g_5,\ g_6,\ g_7\}$
$\newline$
Preferences of boys, most preferred first (only relevant ones): 
$\newline$
$
\begin{array}{|l|l|}
\hline
Pb_1&{g_1}\\
\hline
Pb_2&{g_2,\ g_3,\ g_5,\ g_7}\\
\hline
Pb_3&{g_5,\ g_3}\\
\hline
Pb_4&{g_3,\ g_2}\\
\hline
Pb_5&{g_3,\ g_5,\ g_4}\\
\hline
Pb_6&{g_3,\ g_6}\\
\hline
Pb_7&{g_2,\ g_6,\ g_5}\\
\hline
\end{array}
\newline
$
$\newline$
Preferences of girls, most preferred first (only relevant ones):
$\newline$
$
\begin{array}{|l|l|}
\hline
Pg_1&{b_1}\\
\hline
Pg_2&{b_4,\ b_2},\ b_7\\
\hline
Pg_3&{b_3,\ b_6},\ b_2,\ b_5,\ b_4\\
\hline
Pg_4&{b_5}\\
\hline
Pg_5&{b_7,\ b_2},\ b_5,\ b_3\\
\hline
Pg_6&{b_6,\ b_7}\\
\hline
Pg_7&{b_2}\\
\hline
\end{array}
$
$\newline$
\end{exam}

Running Gale-Shapley algorithm over Example 1 produces the following matching: $b_1-g_1; b_2-g_7; b_3-g_3; b_4-g_2; b_5-g_4; b_6-g_6; b_7-g_5$.
Running Algorithm 2 on the other hand produces the following unique farsightedly stable matching: $b_1-g_1; b_2-g_7; b_3-g_5; b_4-g_2; b_5-g_4; b_6-g_3; b_7-g_6$.

The top-trading-cycles method proposed to be used by Huang in [10] involves taking this outcome and seeking to improve it by materializing trading cycles in some specific order. In particular it materializes first the trading cycle which leave players on that cycle with their top choice (still in play at the time). There are different implementations for the top-trading-cycles method, with an efficient one given in [11]. All implementations however seek to produce the same result. How will the top-trading-cycles method fair on Example 2? Note that there exists a trading cycle in the Gale-Shapley matching for Example 2, $g_2|b_4 \rightarrow g_3|b_3 \rightarrow g_5|b_7 \rightarrow g_2$, where all of $b_4$, $b_3$ and $b_7$ get their first choice (best preferred outcome). As such, any implementation of the method will consider the partial matching $b_7-g_2; b_3-g_5; b_4-g_3$ as part of the final outcome. This is incompatible with the unique farsightedly stable matching produced by Algorithm 2. We could have simply rested our case here, but we believe it is important to illustrate why this difference occurs and why it makes the top-trading-cycles method a poor choice for this instance.

Note for the top-trading-cycles method to be able to materialize even the first trading cycle, it requires the cooperation of $b_2$, $b_5$ and $b_6$ (who could veto the trading cycle). But why would $b_6$ ever cooperate with such a proposal? In fact for him this is a scam. For such an outcome he can always initiate the play $b_6\rightarrow g_3|b_4\rightarrow g_2|b_7\rightarrow g_6$ thus improving things for himself and degrading them for $b_4$ and $b_7$. Note however that $b_4$ and $b_7$ are not required accomplices of this resulting matching. Only $b_2$ and $b_5$ (who are both hopeless men) still need to cooperate for it to materialize. And $b_2$ and $b_5$ are same-off as in either case. Under Assumption 1 they cannot have a preference over which of the other boys to be better off – be it $b_6$ or $b_4$ and $b_7$ – since they still get the same. While this assumption may or may not always hold in practice (for example if $b_2$ is friends with $b_4$) even if they were to veto the farsighted stable matching that would by no means mean that the top-trading-cycles one can materialize: it still requires the cooperation of $b_6$ which can himself then veto it at no cost to him.

Essentially, what was overlooked in [10] is the relative power the players have due to the fact that girls are not “freely tradable” bilaterally between them (as with the original application of the top-trading-cycles method for housing allocation); any trade requires the consent of some of the others. And as we have shown by example, some of these others can get a strictly better outcome for themselves by not cooperating.

This example can also be used to illustrate another phenomenon. If we relax Assumption 2 to allow $b_7$ to make a play which does not farsightedly improve his outcome (in fact it degrades it), namely at $g_5$ then he could take revenge on $b_6$ for not accepting the top-trading-cycles outcome where $b_7$ is better off. This can be argued between players \textit{before} the game actually begins and threats of retaliation could potentially convince some players to make concessions. Threat-making can give raise to interesting outcomes and situations, which are outside the scope of this paper and will be analyzed in a subsequent one. Under Assumption 2, threats cannot exist since they always necessarily involve some self-harmful behavior (when compared to the farsighted coalition stable outcome).

\section{LYING ABOUT PREFERENCES COULD IMPROVE THE FARSIGHTED STABLE MATCHING FOR THE LIER}
For an instance of the Stable Marriage problem with perfect knowledge – where everyone knows the (true) preferences of all the others and of the girls – then the Farsighted Stable Matching is the best anyone can hope for, if we disallow self-harmful threats. 
However, unlike the Gale-Shapley matching where lying can only harm a player’s outcome or leave it unchanged, using deception when negotiating the collusion (before the actual game begins) can bear fruit in the case of the farsighted stable matching. We show this by example. Consider the following instance of the Stable Marriage game.

\begin{exam}
$\newline$
$B=\{b_0,\ b_1,\ b_2,\ b_3,\ b_4,\ b_5\}$
$\newline$
$G=\{g_0,\ g_1,\ g_2,\ g_3,\ g_4,\ g_5\}$
$\newline$
Preferences of boys, most preferred first (only relevant ones): 
$
\begin{array}{|l|l|}
\hline
Pb_0&{g_4},\ g_0\\
\hline
Pb_1&{g_4,\ g_5}\\
\hline
Pb_2&{g_1},{\bf g_3,\ g_4}\\
\hline
Pb_3&{g_1,\ g_2,\ g_3}\\
\hline
Pb_4&{g_2,\ g_1}\\
\hline
Pb_5&{g_5,\ g_2},\ g_4,\ g_1\\
\hline
\end{array}
\newline
$
$\newline$
Preferences of girls, most preferred first (only relevant ones):
$\newline$
$
\begin{array}{|l|l|}
\hline
Pg_0&{-}\\
\hline
Pg_1&{b_4,\ b_5},\ b_3,\ b_2\\
\hline
Pg_2&{b_3,\ b_4},\ b_5\\
\hline
Pg_3&{-}\\
\hline
Pg_4&{b_2,\ b_5},\ b_0,\ b_1\\
\hline
Pg_5&{b_1,\ b_5}\\
\hline
\end{array}
$
$\newline$
Say $b_2$ and only $b_2$ lies about his preference list, swapping $g_3$ and $g_4$.
\end{exam}

The farsighted stable matching over the truthful lists is $b_0-g_0; b_1-g_4; b_2-g_3; b_3-g_1; b_4-g2; b_5-g_5$. However, the farsighted stable matching over the falsified lists is: $b_0-g_0; b_1-g_4; b_2-g_1; b_3-g_3; b_4-g_2; b_5-g_5$. Thus $b_2$ can improve by lying: from his second choice to his best choice.

Lying this way is not permitted by Assumption 2 since it essentially involves the threat of a forbidden play $b_2 \rightarrow g_4$ which is to a worse choice for $b_2$ than he can get under the farsighted stable matching (and also worse than under Gale-Shapley in this case) - under the truthful preference lists of course. Nevertheless, if the other players believed him, he might get away with it. Formally this would be just masked way of making threats of self-harmful actions.

This situation serves to show that players need to be extra circumspect of the preference lists declared by others when deciding to engage in coalitions. Sometimes the preference lists are not declared but are obvious or almost obvious to all players in the game and thus lying cannot succeed. In other cases, like in Student – Medical Program matching, preference lists are submitted to some central authority which then does the matching without them becoming known by others. As such, players wishing to improve by lying may not know precisely \textit{how} to lie given the fact they do not know the preferences of the others.

Ultimately, lying is just a form of threat-making in the hope of convincing others to change their strategy.

\section{Conclusion}
We provided an optimal algorithm for determining the unique farsighted stable matching. The uniqueness of the matching ensures an easy collusion strategy: each player proposes directly to his partner under it. 

We argued by way of example that the matching produced by the top-trading-cycles method, employed in [10] is not always an adequate choice. This has important implications in situations where [10] was employed by some central authority (like potentially in Student – Medical Program matching) as a means to simulate how a Stable Marriage game could be played out, to the advantage of the proposers. In all such cases, [10] should be replaced with Algorithm 2 since the resulting matching could otherwise be unfair to some of the proposers and also unrealistic in the sense that it could not occur in practice when all players are rational.
We have shown that players can improve their outcome under the farsighted stable matching by lying about their preference lists (basically making disguised threats).
We leave open an interesting topic for further research: What happens when threat making is allowed? How can threats be used to gain a better outcome for some boy or group of boys? We will publish some existential results in a subsequent paper.


\section*{Acknowledgment}
No organizations funded the research presented in this paper. The results in sections III of this paper have been available in a rough pre-print form on arXiv, uploaded by Mircea Digulescu there in 2016 [18].
Thanks to Prof. Dr. Andrei Paun, my PhD coordinator for his patient remarks about some of this paper’s content and formatting and for other well-meant advice which lead to a much higher standard of academic writing.
Thanks also to the beautiful people who inspired interest for this topic as these results would never have been discovered (by me) otherwise. 
Finally, thanks also to Jun Wako for making the full-text of \cite{cit21} available on ReasearchGate.com thus enabling us, the Author, to notice that our results had, sadly, been priorly discovered and thus prevent the accidental publication of this paper as containing fresh knowledge. We would like to note however, that we would have preferred it if Dr. Wako was the one to rediscover our results, not the other way around.
Declarations of interest: none.



%

\end{document}